\documentclass[12pt,a4paper]{article}
\usepackage{amsmath,amsfonts}
\newcommand{\e}{\mathrm{e}}
\newcommand{\bea}{\begin{eqnarray}}
\newcommand{\eea}{\end{eqnarray}}

\newcommand{\Tr}{\mathop{\mathrm{Tr}}\nolimits}
\newcommand{\dd}{\mathrm{d}}
\newcommand{\ket}[1]{\left|#1\right\rangle}
\newcommand{\bra}[1]{\left\langle #1\right|}

\newcommand{\pd}{\partial}
\begin{document}

\title{A note on the decay of noncommutative solitons}

\author{Ciprian Acatrinei$^{a,}$\thanks{On leave from: {\it National Institute of
        Nuclear Physics and Engineering  -
        P.O. Box MG-6, 76900 Bucharest, Romania}, e-mail:
        acatrine@physics.uoc.gr;} \hskip 0.2cm and
        Corneliu Sochichiu$^{a,b,}$\thanks{On
        leave from: {\it Bogoliubov Lab Theor Phys, JINR, 141980 Dubna
        Moscow Reg., Russia}, e-mail: Corneliu.Sochichiu@lnf.infn.it}\\
        $^a$ Department of Physics, University of Crete, \\
        P.O. Box 2208, Heraklion, Greece\\
        $^b$ Laboratori Nazionali di Frascati - INFN\\ 
        Via E. Fermi 40 C.P. 13, 
        00044 Frascati, Italy}


\maketitle

\begin{abstract}
We propose an ansatz for the equations of motion of the noncommutative
model of a tachyonic scalar field interacting with a gauge field,
which allows one to find time-dependent solutions describing
decaying solitons. These correspond to the
collapse of lower dimensional branes obtained through tachyon condensation
of unstable brane systems in string theory.
\end{abstract}

\section{Introduction}

Noncommutative field theory was shown to arise as the effective theory
for the massless modes of string theory in
the limit of a
large B-field (see \cite{Seiberg:1999vs}).
The solitonic solutions found in noncommutative field theory
in various limits
(see \cite{Gopakumar:2000zd,Sochichiu:2000rm,Harvey:2000jt,BK} and later works)
are believed to correspond to condensed branes \cite{Sen}, arising from
unstable brane-antibrane configurations \cite{Harvey:2000jt}.
In operator language, these solutions are given by finite rank projectors,
whose Weyl--Moyal forms are localised functions.

Althought dynamical solutions involving noncommutative solitons were
discussed \cite{dyn} the aspects of the collapse process
have still to be 
clarified.\footnote{After the initial version of this note was published
  a number of papers appeared discussing the falling-to-vacuum
  tachyon solutions \cite{tach}.}

The aim of this note is to study the dynamical process of solitonic
collapse in noncommutative theory.
We study the dynamics of rolling-down solutions
in the model of a noncommutative tachyon field $\Phi$ interacting with
a gauge field $A_{\mu}$.
These solutions correspond to unstable noncommutative branes.
They describe the classical decay of noncommutative solitons,
in an infinite time interval.

The plan of the paper is as follows. First we consider the
time-dependent ansatz which reduces the noncommutative field equations
of motion to a chain of separate one-dimensional equations 
which can be implicitely solved. After that we discuss the
implications of the solution, and present some conclusions.

\section{Classical decay}  

The model is given by a $(2+1)$-dimensional action which, in
operator form, reads
\begin{equation}
S=\int\dd t \Tr_{{\cal H}}
\left (
\frac{1}{2}(\nabla_{0}\Phi)^2+\frac{1}{2}[X_i, \Phi]^2 -V(\Phi)
-\frac{1}{4g^2}[X_i,X_j]^2
+\frac{1}{2}(\nabla_0 X_i)^2
\right ), \label{action}
\end{equation}
where $i=1,2$.
The scalar field $\Phi$ and the gauge field $X_{\mu}$ are time-dependent
operators acting on the Hilbert space ${\cal H}$ on which the algebra
\begin{equation}
[x^1,x^2]=i\theta
\end{equation}
is represented.
$X_{i}$ is related to the conventional gauge field $A_{i}$
by $X_{i}=p_{i}+A_{i}$, where
$p_i=\theta^{-1}\epsilon_{ij}x^{j}$.

The equations of motion for the field $\Phi$ are,
\begin{equation}
 \nabla_0^2{\Phi}+[X_{i},[X_{i},\Phi]]+V'(\Phi)=0.
\end{equation}
We consider the case in which the potential $\Phi$ is tachyonic-like,
i.e. it has a local maximum at the origin and local minima at $\Phi=\lambda_i$.

The model admits static solutions (solitons),
\begin{equation}
 \Phi_0=\sum_{i}\lambda_i(1-P_i), \quad [X,\Phi_0]=0,\quad A_0=0, 
 \label{Ansatz1}
\end{equation}
where the operators $P_i$, $P_iP_j=0$ for $i\neq j$,  are projectors to
finite dimensional orthogonal subspaces of the Hilbert space ${\cal H}$.
In Weyl-Moyal language, these solutions are given by functions having
the asymptotics 
$\Phi|_{x \rightarrow \infty}=\lambda_n$.

To show that the solution (\ref{Ansatz1}) is unstable,
consider the following time-dependent ansatz.
The gauge invariance of the model allows one to
choose the operator $\Phi$ to be diagonal.
This partially fixes the gauge.
In the oscillator basis $\{\ket{n}\}$ given by
\begin{equation}
N\ket{n}=n\ket{n}, \quad N=\bar{a}a, \quad
a=\frac{1}{\sqrt{2\theta}}(x^1+ix^2),
\end{equation}
the field $\Phi$ is described by an infinite set of one dimensional
variables
$\Phi_{n}(t)=\bra{n}\Phi(t)\ket{n}$, $n=0,1,2,\dots$
In this basis, the above mentioned static soliton looks as follows:
\begin{equation}
\Phi_n=0,\quad n\leq n_0; \qquad \Phi_n=\lambda, \quad n> n_0.
\label{soliton}
\end{equation}
In what follows,
we will allow dynamics for this finite number of $\Phi_{n}$'s.
In this case, the second equation in the ansatz (\ref{Ansatz1})
is still valid for static $X_{\mu}$.

As a result, the field operator equations of motion split into a set of
decoupled equations
\begin{equation}
\ddot{\Phi}_n+V'(\Phi_n)=0, \qquad 0\leq n \leq n_0,
\end{equation}
with initial conditions for
$\Phi_n$ at $t=0 $ given by (\ref{soliton}),
supplemented with $\dot{\Phi}_n(t=0)=0$ for all $n$.
These equations can be trivially integrated (see any textbook on
classical mechanics),
and the implicit form of the solution is
\begin{equation}
 t(\Phi_n) =\int_{0}^{\Phi_n}\dd\Phi'\frac{1}{\sqrt{V(0)-V(\Phi')}},
\qquad n\leq n_0,
\end{equation}
and $\Phi_n=\lambda$, for $n>n_0$.
If one integrates between a local maximum ($\Phi=0$)
and an adjacent minimum ($\Phi=\lambda$) of the potential,
the above integral gives the time of fall, which diverges logarithmically
as $\Phi$ approaches the origin.

Thus  classically the soliton will decay into an infinite amount of time.
This is the standard situation for unstable equilibrium points
in classical mechanics.
However, quantum fluctuations will kick the soliton out of the origin
in a finite time.

In particular, if one limits oneself to the lowest terms of the
tachyonic potential, $V(\Phi)=V(0)-\frac12 m^2\Phi^2+g\Phi^4$,
the classical solution is
\begin{equation}\label{cl_sol}
\Phi_n(t)=\frac{m}{\sqrt{g}}\frac{2\e^{-mt}}{1+\e^{-2mt}},
\qquad n\leq n_0,
\end{equation}
which satisfies $\Phi(t=-\infty)=0$ and
$\Phi(t=0)=\frac{m}{\sqrt{g}}=\lambda$ for $n\leq n_0$.
In operator form this solution looks like
\begin{equation}
\Phi(t)=\frac{m}{\sqrt{g}}\frac{2\e^{-mt}}{1+\e^{-2mt}}P_{n_0},
\end{equation}
where $P_{n_0}$ is the projector operator to the first
$n_0+1$ states $P_{n_0}=\sum_{n=0}^{n_0}\ket{n}\bra{n}$.

We also note the existence of Euclidean-time (instantonic) solutions of the
equations of motion, which interpolate between two degenerate vacua:
\begin{equation}
\Phi(\tau): \quad\tau=\int_{\lambda_i}^{\Phi}\frac{d\tau}{\sqrt{V(\Phi)}},
\quad \tau=it.
\end{equation}
In the real time approach they correspond to tunnelling between the vacua.

%
%

Let us make a remark.
We considered time-dependent solutions with static gauge field configuration,
leaving aside the possible evolution of the gauge fields.
There are indications however that those may decay as well. 
Indeed, the \emph{most general\/} gauge field background
satisfying (\ref{Ansatz1}) is given \cite{Sochichiu:2000rm}
by the block operator,
\begin{equation}\label{x1}
  X_\mu=
  \begin{pmatrix}
  X^{(n)}_\mu & 0 \\
  0 & X^{(\infty)}_\mu
  \end{pmatrix},
\end{equation}
where $X^{(n)}_\mu$ are $n\times n$ Hermitian matrices,
while $X^{(\infty)}_\mu$
are (infinite-dimensional) Hermitian operators.
Both of them satisfy 
$$ 
\ddot{X}_\mu-\frac{1}{g}[X_\nu,[X_\nu,X_\mu]]=0,
$$
where $X$ stands for either $X^{(n)}$ or $X^{(\infty)}$.
In particular one can choose a solution where
$X^{(n)}=0$ and $X^{(\infty)}$ are
some new $p_\mu'$ satisfying the commutation relation on the infinite
dimenstional subspace $[p_\mu',p_\nu']=i
\theta^{-1}_{\mu\nu}$.

The finite dimensional part of the gauge field given by $X^{(n)}$ possesses
\emph{only\/} static solutions satisfying
$[X^{(n)}_\mu,X^{(n)}_\nu]=0$, which are \emph{stable} since they have
zero energy (density). So the only trouble may come from the
infinite-dimensional sector $X_\mu^{(\infty)}$ which, beyond the
commuting solutions of the above type, may also have solutions with
c-number commutator. These
gauge field configurations possess finite energy density proportional
to the square of the inverse noncommutativity parameter
$(\theta^{-1}_{\mu\nu})^2$, and therefore potentially may decay to a
commutative solution with $\theta^{-1}=0$, or equivalently
\begin{equation}\label{commutingX}
  [X^{(\infty)}_\mu,X^{(\infty)}_\nu]=0,
\end{equation}
which has zero energy density. For a description of the noncommutative
model around the background (\ref{commutingX}) we refer the reader to
the paper \cite{Sochichiu:2000ud}. 

The solution found in this paper gives one channel of decay.
As it often happens with unstable systems, there could be other channels
on which the system classically slides down.
In the quantum theory these channels would be assigned different probabilities.
Such an analysis analysis, however, goes beyond the scope of the present work.
We just mention that, from another point of view, there are indications of
instabilities of purely bosonic noncommutative Yang--Mills models at
one-loop level \cite{instab}.

\section{Radiation}

Finally, we discuss the possible radiation during the decaying process.
We consider for that matter the general tachyonic action (\ref{action})
where $\nabla_0 =\pd_0 + [A_0,\cdot]$, and all fields are Hilbert space
operators. 

The solution of the general equations of motion,
\begin{align}\label{phi}
  &\nabla_0^2 \Phi +[X_i,[X_i,\Phi]]+V'(\Phi)=0,\\
  &\nabla_0^2X_i+\frac{1}{g^2}[X_k,[X_k,X_i]]+[\Phi,[X_i,\Phi]]=0,\\
  \label{x} 
  &[X_i,\dot{X}_i]+[\Phi,\nabla_0\Phi]=0,
\end{align} 
is given by the ansatz (in diagonal gauge),
\begin{equation}
  \ddot{\phi}+V'(\phi)=0,\qquad [Y_i,\phi]=[A_0,\phi]=0.
\end{equation} 

Let us label the components of the background solution as follows
$\phi_a=\phi(t)\neq \lambda$, $a=1,\dots,N$  and $\phi_n=\lambda$,
$n=0,1,\dots$ where $\lambda$ is the vacuum of the tachyonic
potential. In above index $a$ spans the solitonic subspace of the
Hilbert space while $n$ spans its infinite dimensional completion.

According to these notations the gauge field is split in two blocks
corresponding to respectively tachyonic subspace and its completion, each of
these blocks satisfying separately,
\begin{equation}\label{em}
  [Y^{(A)}_k,[Y^{(A)}_k,Y^{(A)}_i]]=0,
\end{equation}
where capital $A$ denotes respectively blocks corresponding either to
solitonic space or to its completion.

Fields in the solution are assumed to have the following block
structure,
\begin{equation}\label{blocks}
 \phi=
 \begin{pmatrix}
 \phi & 0 \\
 0    & \lambda
 \end{pmatrix}, \qquad
 Y_i=
 \begin{pmatrix}
 Y^{(0)}_i&0\\
 0 & Y'_i
 \end{pmatrix}.
\end{equation}

From \eqref{em}, the solitonic block should be diagonal (in the
basis in which the tachyon is diagonal), while the infinite dimensional
block in the completion space could be choosen e.g. to satisfy
\begin{equation}\label{yprime}
  [Y'_i,Y'_j]=B_{ij}, \quad \det B\neq 0,
\end{equation}
giving rise to a noncommutative space by itself.

We now show that the decaying tachyonic soliton is a source for 
various particles living in the space generated by $Y'$ satisfying \eqref{yprime}. 
To see this, consider the fluctuation fields around the solution,
\begin{align}
  & \Phi=\phi+u,\\
  & X_i= Y_i+a_i,\\
  & A_0= a_0.
\end{align}
Inserting the above into the action yields after some algebra, 
up to second order in fluctuations,
\begin{multline}
  S[u,a]= S[\phi,Y]+\\
  \frac12 (\nabla_0 u)^2-\frac12 V''(\phi)u^2-\frac12[Y_i,u]^2+\\
  \frac{1}{2g^2}(\nabla_0 a_\mu)^2-\frac{1}{4g^2}[Y_i,a_\mu]^2+\\
  \frac12 [a_\mu,\phi]^2+\dot{\phi}[a_0,u]-[Y_i,u][a_i,\phi].
\end{multline}
The lines two and three in the above equation give the free propagators of
respectively $u$ and $a_\mu$ fields while the last line
provides the interaction vertices. We also used the gauge fixing condition
$ \nabla_0 a_0-[Y_i,a_i]=0.$

If we split the fields $u$ and $a_\mu$ into blocks according to the
block structure \eqref{blocks},
\begin{equation}
 u=
 \begin{pmatrix}
 c       & v \\
 \bar{v} & u'
 \end{pmatrix},\qquad 
 a_\mu=
 \begin{pmatrix}
 c_\mu  & b_\mu \\
 \bar{b}_\mu & a'_\mu
 \end{pmatrix}.
 \end{equation}
the decaying soliton field is a source for
the particles $b$, $\bar{b}$, $v$ and $\bar{v}$, as follows,  
\begin{enumerate}
\item $\frac12 [a_\mu,\phi]^2\to [b_\mu,\phi][\bar{b}_\mu,\phi]$ --- a
  pair of charged photons 
  $b_\mu\bar{b}_\mu$.
\item $\dot{\phi}[a_0,u]+[Y_i,u][a_i,\phi]\to
  \dot{\phi}[b_0,\phi]+[Y_i,v][\bar{b}_i,\phi]+$h.c. --- a charged
  photon + the scalar with opposite charge, where h.c. stands for
  Hermitian conjugate.  
\end{enumerate}

We see that at tree level the decay process produces only particles in
the fundamental representation of the unbroken gauge
group\footnote{More precisely: in the bi-fundamental representation of
  each factor.}. Because of gauge invariance emission is possible for
only totally neutral pairs.

\vskip 0.3cm

\section{Conclusion}
Let us conclude with the following:

\begin{itemize}

\item In string theory framework, noncommutative solitons correspond to the
D(p-2)-branes which result from condensation of unstable D(p)-brane systems.
As our analysis shows, the D(p-2)-brane appears to be unstable as well,
decaying into nothing.
\item In their classical rolling down,
solitons will be reflected back and reach
their initial values (in infinite time).
As we have shown, however, the decaying brane can emit definite types of
charged particles and antiparticles. 
In this case, the field will not reach its initial value,
but will oscillate, with decreasing amplitude, around the vacuum.
Then noncommutative solitons could have cosmological applications,
playing the role of the inflaton field.

\item For a general noncommutative potential the solitons are unstable
iff there are states in the Hilbert space for which the noncommutative field
$\Phi$ takes values corresponding to unstable points of the potential (as a
commutative function). In the case of solitons living at local minima, they are
unstable due to tunnelling to the global minimum.

\end{itemize}
\subsection*{Acknowledgements}
We thank the HEP theory group of the University of Crete for hospitality and
stimulating discussions. The work of C.A. was supported by a MURST (Italy)
fellowship. The work of C.S. was partially supported by
RFBR grant \#99-01-00190,
INTAS grants \#1A-262 and \#99 0590, Scientific School support
grant \# 00-15-96046 and NATO fellowship program.


\end{document}